# What, When and Where of petitions submitted to the UK Government during a time of chaos

Bertie Vidgen and Taha Yasseri*
Oxford Internet Institute, University of Oxford, Oxford, UK
Alan Turing Institute, London, UK

*Corresponding Author: taha.yasseri@oii.ox.ac.uk

## Abstract

In times marked by political turbulence and uncertainty, as well as increasing divisiveness and hyperpartisanship, Governments need to use every tool at their disposal to understand and respond to the concerns of their citizens. We study issues raised by the UK public to the Government during 2015-2017 (surrounding the UK EU-membership referendum), mining public opinion from a dataset of 10,950 petitions (representing 30.5 million signatures). We extract the main issues with a ground-up natural language processing (NLP) method, latent Dirichlet allocation (LDA). We then investigate their temporal dynamics and geographic features. We show that whilst the popularity of some issues is stable across the two years, others are highly influenced by external events, such as the referendum in June 2016. We also study the relationship between petitions' issues and where their signatories are geographically located. We show that some issues receive support from across the whole country but others are far more local. We then identify six distinct clusters of constituencies based on the issues which constituents sign. Finally, we validate our approach by comparing the petitions' issues with the top issues reported in Ipsos MORI survey data. These results show the huge power of computationally analyzing petitions to understand not only *what* issues citizens are concerned about but also *when* and from *where*.

**Keywords**: petition, political participation, Brexit, opinion mining, government

## 1 Introduction

Contemporary politics is marked by increasing turbulence (Margetts et al. 2015), from surprise election results, such as Theresa May's slender majority in 2017, to seismic political shifts, such as the Brexit vote in 2016, and party schisms, such as the 11-MP breakaway to Change UK in 2019. Uncertainty has increased as traditional markers of political affiliation, such as party membership, have declined in importance, and politics is seen to have become dominated by identity- and issue- based activism (van Biezen and Poguntke 2014; Saunders 2014; Schumacher and Giger 2017). At the same time, governments are perceived to have failed in listening and responding to the concerns of the public, precipitating a rise in populism and anti-elitism. Populist parties, which allege that mainstream governments are disconnected from the people, self-interested and, in some cases, undemocratic, have become increasingly popular across Europe over the last decade (Cleen and Stavrakakis 2017; Oliver and Rahn 2016; Webb and Bale 2014). In this unpredictable and contentious political environment, there is an even greater need for governments to understand the concerns of the public, and to reflect this in their agenda, discourse, and policies.
Signing a petition is one of the few ways in which citizens can easily and legally raise issues in between elections (Hough 2012; Lindner 2011; Stewart, Cuddy, and Silongan 2013), and can be considered a micro-act' of unconventional political participation (Margetts et al. 2015). Petitions have been described as a 'tool for the voicing of grievances' (Melo and Stockemer 2014) and they are an excellent source of data for mining the concerns of the public directed to the UK government for three reasons. First, the government has its own petition website. The petitions hosted on there are explicitly directed to the government and as such they are a very explicit form of unconventional political participation. Second, online petitions are widely used; Hansard reports that 28% of the public have created or signed an e-petition (The Hansard Society 2018) and Dutton and Blank report that over 1 in 3 people would consider doing so (Dutton and Blank 2013). Third, petition signing is connected to people's broader engagement with issue-based politics; as Gibson and Cantijoch write, 'one can more easily move from signing an e-petition to contacting a politician or volunteering to help a party' (Gibson and Cantijoch 2013, 714).

Much existing research on petitions has focused on the dynamics of petition signing (Böttcher, Woolley-Meza, and Brockmann 2017; Hale et al. 2018; Yasseri, Hale, and Margetts 2017) and on understanding the political significance of petitioning (Jungherr and Jürgens 2010; Lindner 2011; Wright 2016). Limited research has investigated the thematic content of petitions. Puschmann et al. analyse the content of petitions submitted to the German Bundestag and find different policy issues attract signatures from different types of signatories. Some issues, like 'Labour' and 'Transport', are dominated by signatories who have signed many petitions whilst others, like 'Science', are dominated by 'sporadic' signatories (Puschmann, Bastos, and Schmidt 2017). Hagen et al. report a similar result studying petitions submitted to the USA government. They also find that in some cases the number of signatures received by petitions can be linked to external factors. For instance, the prevalence of petitions associated with the issues "Japan" is linked to fluctuations in relevant Google search terms but this is not the case for the issue 'Animal' (Hagen et al. 2015).

Panagiotopoulous et al. look at the relationship between the number of signatures received by petitions, their topic and whether they have dedicated user community pages on Facebook. They find that petitions with large numbers of signatures do not necessarily have large Facebook communities, and that this is linked to the petitions' topic popularity (Panagiotopoulos et al. 2011). Clark et al. study the location of signatories to petitions submitted to the UK government which crossed the threshold for a government response (>10,000 signatures). They identify four classes of users; Domestic Liberals, International Liberals, Nostalgic Brits and Rural Concerns. Each class is associated with (i) thematically different petitions, such as 'environmental protection' or 'the EU referendum', and (ii) different substantive positions, such as being for or against the UK leaving the EU. Clark et al. provide evidence of a strong relationship between the geographic location of signatories and petitions' content, but only on a small dataset (Clark, Lomax, and Morris 2017).

Previous work shows that petitions are a useful source of data for understanding not only *what* issues people care about but also *when* and from *where*. However, no study has integrated analysis of these dynamics and considered all petitions submitted to a national government, including those which receive few signatures. A considerable challenge in this domain is the sheer volume of petitions which are created and signed, and the wide variety of issues and outlooks they cover. This makes it difficult to read, analyze and summarize them in a timely manner (Grimmer and Stewart 2013) both for researchers and Government. Hence, apart from the very few successful petitions that receive a formal response, the rest turn into digital dust. In this paper we respond to this gap in existing research and computationally analyze all petitions submitted onto, and are publicly available from, the UK government's petition platform during 2015-2017.

# 2 Results

During the 2015-2017 parliament 31,173 petitions were submitted to the UK Government, of which 10,950 petitions were accepted onto the platform. They collectively received 31.5 million signatures. 486 petitions reached the threshold for a response from the government (10,000 signatures) and, of these, 65 petitions reached the threshold for a debate by a House of Commons Select Committee (100,000 signatures). The vast majority of petitions receive very few signatures – 64% of petitions (7,034) received fewer than one hundred signatures – whilst a small number receive many; the top 10 petitions received 30% of all signatures (9,468,477). The most successful petition, which called for a second referendum on leaving the EU, received 4.15 million signatures. Table 1 shows the top ten petitions launched during the period, ranked by number of signatures. Noticeably, three of the petitions are about Donald Trump, two expressing opposition and one expressing support.

Table 1: Top ten petitions launched during the 2015-2017 parliament, ranked by number of signatures.

| **Action called for** | **Number of signatures** | **Date of creation** |
|---|---|---|
| EU Referendum Rules triggering a 2nd EU Referendum | 4,150,262 | 23rd May 2016 |
| Prevent Donald Trump from making a State Visit to the United Kingdom | 1,863,707 | 9th Nov 2016 |
| Give the Meningitis B vaccine to ALL children, not just newborn babies | 823,349 | 9th Sep 2015 |
| Block Donald J Trump from UK entry | 586,930 | 28th Nov 2015 |
| Stop all immigration and close the UK borders until ISIS is defeated | 463,501 | 5th Sep 2015 |
| Accept more asylum seekers and increase support for refugee migrants in the UK | 450,287 | 10th Aug 2015 |
| Consider a vote of No Confidence in Jeremy Hunt, Health Secretary | 339,925 | 10th Feb 2016 |
| Donald Trump should make a State Visit to the United Kingdom | 317,542 | 29th Jan 2017 |
| Make the production, sale and use of cannabis legal | 236,995 | 20th Jul 2015 |
| Stop spending a fixed 0.7 per cent slice of our national wealth on Foreign Aid | 235,979 | 24th Mar 2016 |

Figure 1 shows the empirical complementary cumulative distribution function for the observed number of signatures per petition, with indicators at the government thresholds of 10,000 and 100,000 signatures. Up to 10,000 signatures the data fits closely to a power law distribution with a minimum value of 10 and an estimated exponent of 1.42 (Clauset, Shalizi, and Newman 2009). However, after this first threshold is reached the probability of a petition receiving a given number of signatures is lower than that of the same power law distribution. A further, steeper, divergence is observed at the second threshold of 100,000 signatures. These two divergences suggest that the government's thresholds markedly influence the behavior of signatories on the petition platform. Potentially, once a government threshold is reached, petition creators campaign less actively to attract signatures or signatories are less motivated to sign a petition as they believe it has achieved 'success'. There is also a fourth group of petitions at the right end of the distribution, which appear to deviate in the opposite direction.

Those are mainly petitions that received considerable media attention after crossing the Governments' thresholds.

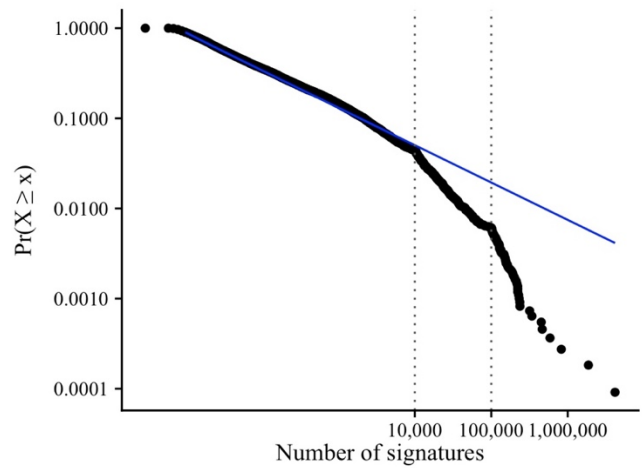

Figure 1: Complementary Cumulative Distribution Function for the number of signatures per petition. The red line is a fitted power, vertical dotted grey lines show government response thresholds.

## 2.1 Issues

Due to the large number of petitions (n = 10,950), the corpus cannot be easily read and annotated manually. We use an unsupervised NLP algorithm, Latent Dirichlet Allocation (LDA) (Blei et al. 2003; Blei, Carin, and Dunson 2012), to extract topics from all of the free text fields filled in by petition creators. We fit a model with ten topics, which are distributions over the entire vocabulary, and name each of them as issues (see  ).[1] The ten issues, and the words most strongly associated with them, are shown in Table 2 in descending order. The issues are well separated (i.e. the top words do not overlap). They relate to recognizable political concerns, from mainstream issues, such as 'International Affairs' and 'Law & Order', to more niche issues, such as 'Animals & the Environment' and 'Driving'. Petitions which are associated with the same issue can express very different types of sentiment and ideology and call for very different actions to be undertaken. For instance, both of the petitions shown below have a very high score of being related to the 'School' issue (0.96 and 0.95 respectively). The first petition supports a career-focused approach to education:

> "The UK needs a modern integral education based on the talents and abilities of students; so that when they finish secondary they know three languages and have a defined job."

In contrast, the second petition supports a vocational approach to education:

> "In state education here in the U.K our youth have next to no chances to properly further their vocational skills, e.g. dance! Many schools not offering a thing in some of these areas that could potentially be the child's gift in life!"

---

[1] Topic refers to the distributions over words generated by the LDA topic model. Issue refers to the named topics, which we analyse.

Even though very different views are expressed in the petitions, they are both highly loaded on the School issue.

Table 2: Top six terms for the ten issues

| Issue name | Word 1 | Word 2 | Word 3 | Word 4 | Word 5 | Word 6 |
|---|---|---|---|---|---|---|
| International affairs | British | Govern | Country | Nation | World | Citizen |
| Democracy & the EU | Vote | Referendum | Govern | Parliament | Leave | Will |
| Law & Order | Law | Police | Act | Public | Protect | Crime |
| School | School | Children | Student | Education | Year | Young |
| Driving | Road | Car | Use | Driver | Drive | Vehicle |
| Family | Children | Child | Parent | People | Family | Need |
| Work & Pay | Pay | Tax | Work | Year | Cost | Money |
| Animals & the Environment | Dog | Animal | Ban | Use | Food | Can |
| Healthcare | NHS | Health | People | Care | Mental | Need |
| Local Government | Govern | Housing | Local | Will | Council | Fund |

*Prevalence of Issues*

We identify the most prevalent issues (i) based on the number of petitions and (ii) weighted by the number of signatures each petition receives. To measure prevalence in terms of the number of petitions we sum all of the petition-specific topic distributions, giving each petition a weight of 1. To weight prevalence by the number of signatures, we multiply each petition's topic distribution by the number of signatures the petition receives. Both analyses are shown in Figure 2. The unweighted distribution petitions over topics is broadly uniform with a weak right skew. In contrast, the distribution of petitions weighted by signatures has a strong right skew. In most existing research into petitions, the primary unit of analysis is the petition itself. Yet the striking discrepancy between the first two panels in Figure 2 demonstrates the importance of explicitly modelling the number of signatures rather than just the number of petitions. Otherwise, too much attention is paid to issues which appear in many petitions but attract few signatures, which can distort analysis. Our counter-intuitive approach is to abstract away from the petition itself and focus instead on (i) petitions' issues and (ii) the number of signatures petitions receive. This better captures the political act of participation we are primarily interested in, which is signing petitions rather than creating them.

Based on the number of signatures, the most prevalent issue is 'Democracy & the EU' (7.5 million signatures), followed by 'International Affairs' (5.8 million signatures) and 'Healthcare' (3.1 million signatures). The large number of signatures 'Democracy & the EU' receives is expected given that it contains many petitions related to the EU Referendum, which was a key political issue in the UK during 2015 to 2017. Similarly, there were many important and widely reported foreign affairs events, such as conflict between NATO and ISIS in the Middle East. Six of the ten issues receive between two and three million signatures. 'School' and 'Family' receive 1.9 million and 2.1 million signatures respectively (2[nd] and 3[rd] fewest signatures), which is surprising given that these receive considerable media attention and are often viewed as key concerns in society. 'Driving' has the fewest signatures (1 million), which is expected given that it is a fairly niche issue.

The right panel of Figure 2 shows the posterior probability that petitions assigned to each issue receive 10,000 signatures or more. It is notable that the issues which are associated with the most signatures are not necessarily the most likely to be successful. Petitions relating to 'Democracy & the EU' receive the most signatures overall but are only the fourth most likely to receive 10,000 signatures or more and so receive a response from the government. This is likely because there are a few 'super petitions' for these issues which attract millions of signatures – but this still only equates to one successful petition. Overall, there is relatively little variation in the probability of success, which ranges from 0.031 to 0.056. This supports previous research which indicates that the content of petitions is not a significant factor in determining whether they are successful (Margetts et al. 2015).

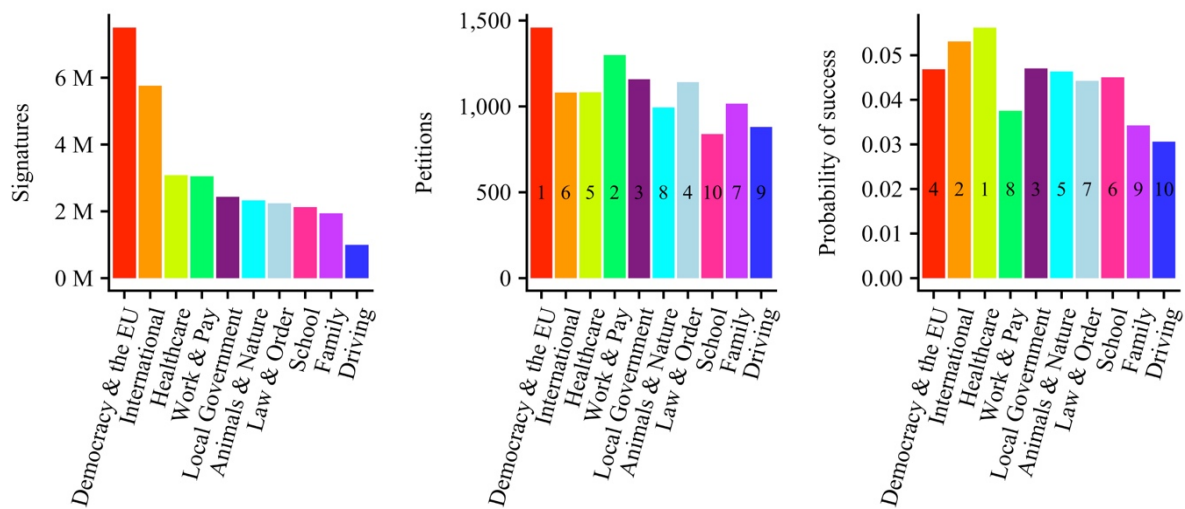

Figure 2, The prevalence and success of issues. Left: the distribution of signatures over issues. Middle: the distribution of petitions over issues. Right: the probability that petitions assigned to each issue will receive 10,000 signatures or more. The numbers show the issues' ranked position.

*Relationships between issues*

To better understand the connections between issues, and how petitions join different issues together, we study (i) the co-occurrence of issues within same petitions and (ii) how similar issues are in terms of their words distributions. In both cases we measure the relationships between issues using cosine similarity. Figure 3 (Left) shows a network of issue co-occurrence within petitions. Issues are related weakly in terms of how they co-occur within petitions. Cosine values are fairly low, with an average of 0.11 and a range of 0.05 (between 'Driving' and 'Democracy & the EU) to 0.2 (between 'Law & Order' and 'Family'). This is likely affected by the fact that most petitions are dominated by a single issue; on average, petitions' most probable issue has a probability of 0.62. This is understandable given the relatively small amount of space petition creators are given to write about their petition, and the fact that petitions are a single-issue based form of political participation. There are considerably stronger links between issues in terms of how similar their words distributions are. Calculated on a pairwise basis, the average cosine similarity is 0.27. The greatest similarity is between School and Family (cosine = 0.44) and the weakest is between 'Democracy' and 'Driving' (cosine = 0.18). The similarity of issues based on topic word distributions is presented as a network in the right panel of Figure 3.

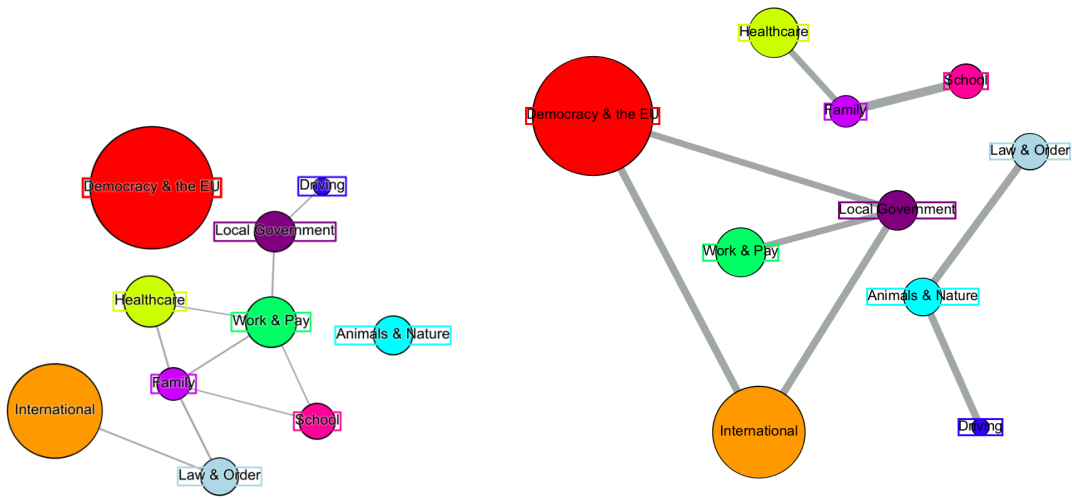

Figure 3: Network of issues based on their similarity. Left: Network of issue co-occurrence within petitions. Edges are weighted by the cosine similarity of topics co-occurrence. Right: Network of topic similarity based on word distribution. Edges are weighted by the cosine similarity of word-topic distributions. Nodes are weighted by the number of signatures each topic receives. Only the top 20% strongest edges are shown.

## 2.2 The dynamics of issues over time

Issues show different temporal dynamics, whereby some exhibit large fluctuations over time and others exhibit minor fluctuations in prevalence and popularity.[2] The dataset contains the day on which petitions are created and the total number of signatures they receive. We do not have the daily counts of signatures received by each petition. However, past research has shown that the vast majority of signatures come within the first few days after when a petition is created (Yasseri, Hale, and Margetts 2017). Hence, we use the creation date as a proxy for the time at which each petition receives signatures.

**Error! Reference source not found.** shows the signatures received by issues plotted over time. The upper left panel shows the values on a linear scale. The issue 'Democracy & the EU' has a very noticeable spike in May 2016 due to a highly popular petition which called for the EU referendum vote to be repeated. Similarly, the issue 'International Affairs' has large spikes in November 2016 due to a highly-publicized petition which called for Donald Trump to be banned from the UK (see Table 1). Most other issues remain broadly stable over the time period with only small fluctuations in popularity. This suggests that whilst some issues have a relatively constant presence, others vary more as their signatures are driven by exogenous events. A limitation of this graph is that the large number of signatures received by the 'Democracy & the EU' issue makes it difficult to observe fluctuations in other issues. Accordingly, in the upper right panel the same data is plotted with a logarithmic scale (base 10). In the four remaining panels we plot the number of signatures received by issues, smoothed with various time windows.

---

[2] Prevalence refers to the overall number of signatures received by each petition or topic. Popularity refers to the relative interest in each petition or topic.

The large number of signatures received by 'Democracy & the EU' in May 2017 remains visible across all four smoothing time windows, indicating that this fluctuation is signal rather than noise. Similarly, for 'International Affairs' at least two peaks can be observed in all four plots; one in late 2015 (driven primarily by a very popular anti-Trump petition) and the other in early 2017 (also driven primarily by a very popular anti-Trump petition). This suggests that for both these issues signatures are primarily driven by exogenous events. In contrast, for the other eight issues the fluctuations decrease noticeably as the time window increases, which suggests that signatures are broadly stable and not driven by external events.

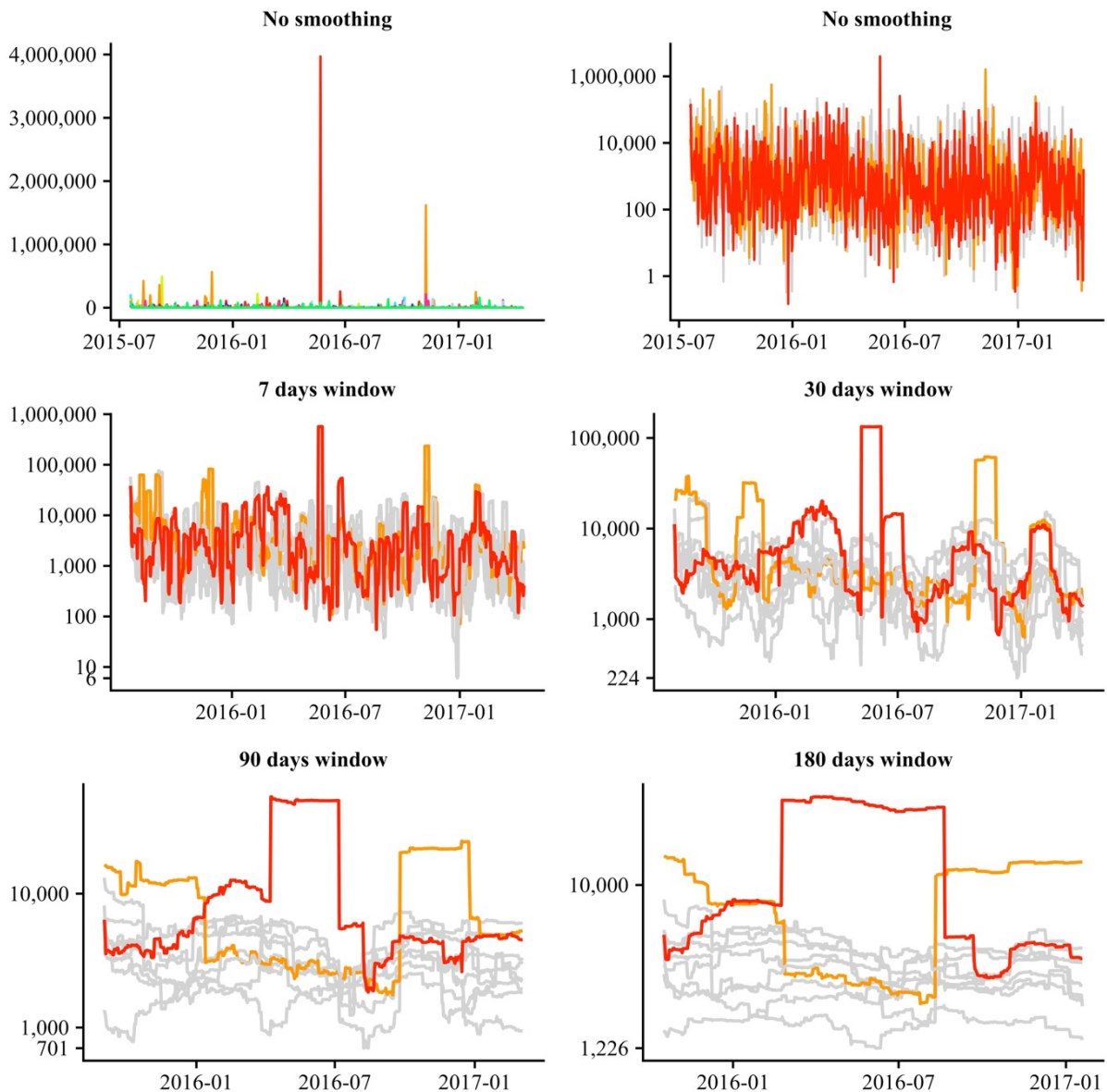

Figure 4: Signatures received by issues over time. The upper left panel has a linear scale and the upper right panel has a logarithmic scale (base 10). The remaining four panels show the data smoothed with time windows of one week, one month, three mon

*Volatility: entropic change*

The previous section shows that the prevalence of issues changes considerably over time. To examine this systematically, we calculate normalized Shannon information entropy on the distribution of signatures over issues. Entropy is a measure of the level of disorder within a system, widely used to study the inequality of distributions (Shannon 1948). We use a one-week time window to calculate entropy as we observe a strong weekly pattern in the volume of signatures.

The left panel of **Error! Reference source not found.** shows the normalized entropy plotted over time. The range of values is between 0.114 and 0.533, and the mean is 0.438. The left panel in **Error! Reference source not found.** shows the daily percentage change in entropy. We define substantial changes in entropy as daily percentage changes which are more than three standard deviations from the mean percentage change (0.8%) as this equates to a 0.01 significance level with normally distributed data (Vidgen and Yasseri 2016). This is shown by the grey dotted lines on the right panel. The grey dotted lines in the left panel show dates which fall outside of the 3 standard deviation range. Overall, nine dates are identified where the entropy changes substantially, of which 6 are due to increases in entropy and 3 are due to decreases. A noticeable period of entropic change occurs from the 6th November 2016 to the 30th December 2016, when 5 days out of 55 record substantial changes. This is surprising given that on only one day during this period is there a noticeable peak in the prevalence of a single issue (on 9th November 2016 when an anti-Trump petition was launched), as shown in **Error! Reference source not found.**. This demonstrates that significant changes in entropy can occur even when the absolute number of signatures is quite small and changes are not immediately apparent from qualitative analysis.

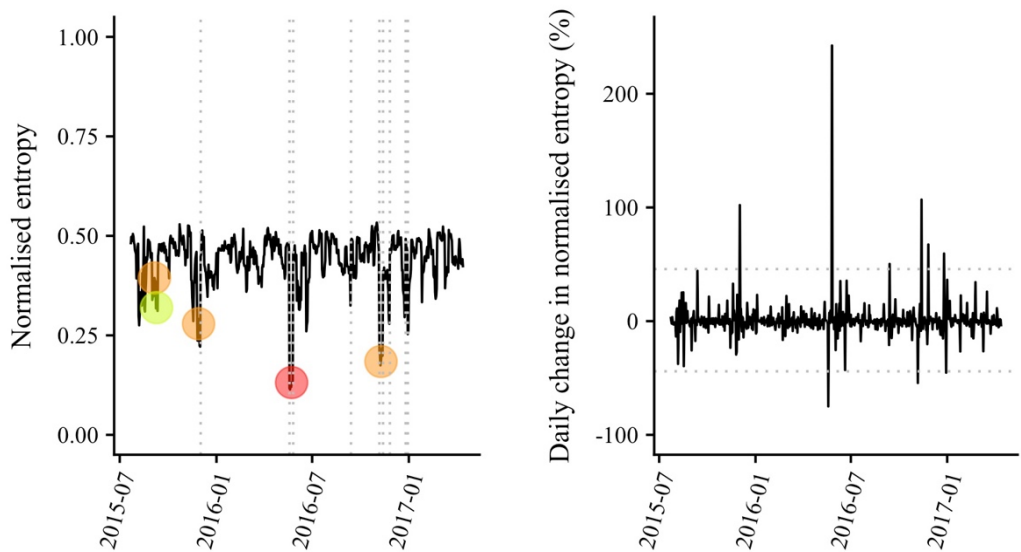

Figure 5: Entropic change over time. Left: Normalized Shannon entropy for the distribution of signatures over issues. Right: Daily percentage change in normalized entropy.

### 2.3 The geography of issues

Different geographic areas sign petitions associated with different to issues. Our data contains the number of signatures from each parliamentary constituency (n = 650) for each petition. The left and middle panels of Figure 6 show the distribution of signatures per constituency in total and distribution of signatures per electorate in each constituency. The mean number of signatures per constituency is 46,800 and the mean number of signatures per electorate is 0.65. The plots are similarly distributed, both with a very moderate right skew. The constituencies with the most signatures are Bristol West (n=135,499), Brighton Pavilion (n=120,453) and Bethnal Green and Bow (n=106,218). This matches with the constituencies with the most signatures per electorate, which are Brighton Pavilion (n=1.6), Bristol West (n=1.46) and Hornsey and Green Wood (n=1.3). The number of signatures per constituency and the number of signatures per electorate is also mapped in Figure 7 for the 632 constituencies in Great Britain.

The right panel in Figure 6 shows the number of signatures per constituency plotted against the size of the constituency's electorate at the 2017 General Election. We log the transform the data and fit a linear regression model. The model has an R-squared of 0.39 and an exponent of 1.47. This indicates a strong scaling relationship, whereby constituencies with larger electorates sign comparatively more petitions per person than those with smaller electorates. To validate the result, we re-run the analysis using binned data, and also calculate a super linear relationship (exponent = 1.32). This indicates that constituents' average engagement with petitions increase as the size of constituency that they live in increases.

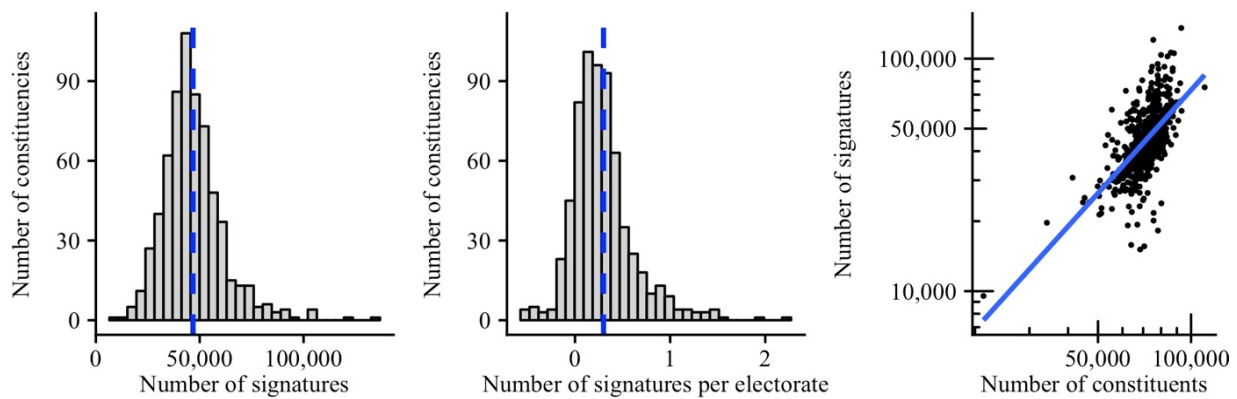

Figure 6: Signatures per constituency. Left: Histogram of the total number of signatures per constituency. Middle: Histogram with number of signatures per electorate. Right: Number of signatures vs number of constituents.

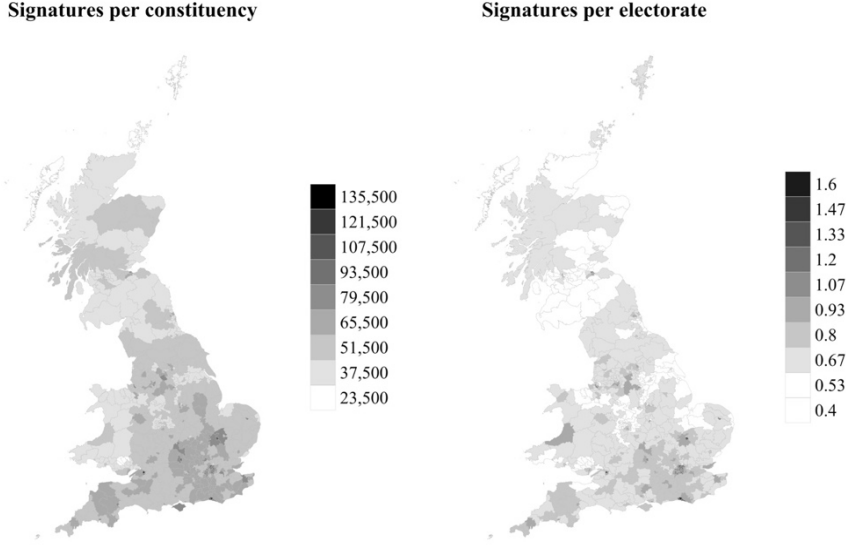

Figure 7: Number of total signatures per constituency and per electorate.

*Geographical prevalence of issues*

The previous section demonstrates that the level of petition signing varies considerably across different constituencies, and that this has a super-linear relationship with the size of each electorate's constituency. We provide more granular insight in this section by investigating how the number of signatures received by each issue varies by constituency. The results of this analysis are visualized in Figure 8. For each constituency, the percentage of signatures given to each issue is calculated. This controls for the different total number of signatures in each constituency (Figure 7). However, the percentages give limited insight on their own as nearly all constituencies give most of their signatures to the most popular issues. To account for this, we calculate Z-scores for each issue, based on the percentage of signatures given by each constituency. This enables us to compare the *relative* importance of each issue within each constituency. This is shown in Eq. (1) and (2) below, where c is the constituency, *s* is the number of signatures, *i* is the issue and *z* is the score we assign.

$$sp_{ci} = \frac{s_{ci}}{s_c}, \qquad \mu_i = \frac{\sum_{c=1}^{C} sp_{ci}}{C}, \qquad \sigma_i = \sqrt{\frac{\sum_{c=1}^{C}(sp_{ci} - \mu_i)^2}{n-1}}$$

(1)

$$z_{ci} = \frac{sp_{ci} - \mu_i}{\sigma_i}$$

(2)

Several issues can be identified as National issues, including, 'Law & Order' and 'Work & Pay'. The number of signatures given to these issues is more uniform than for other issues. They attract support from many different parts of the country, and the variations do not follow a discernible pattern. In contrast, other issues are highly regional. 'Driving' is highly important for a small set of constituencies in the South East but less important elsewhere. 'Animals & Nature' is particularly notable; urban areas, including London, the Midlands and northern cities assign very little attention to the issue, rural constituencies assign more attention, and areas of natural beauty which are far from urban centers, including Cornwall, West Wales and North

Scotland, assign it the most importance. Scotland has distinctive dynamics for several issues; 'International' is broadly a national issue but is especially important in Scotland. Conversely, 'Local Government' and 'School' have far fewer signatures in Scotland. 'School' has very little importance in Scotland, which suggests that differences in Schooling policy between devolved governments has had an impact on public attitudes.

The maps also indicate that London has distinctive petition signing habits. It assigns far less importance to 'Local Government', 'Healthcare', and 'Family', and far more importance to 'Democracy & the EU' and 'International'. This reflects a broader pattern where, in general, petition signing habits vary between rural and urban constituencies. Rural constituencies tend to petition about traditional domestic political issues whilst urban areas are more concerned about ideological issues, such as 'Democracy & the EU'.

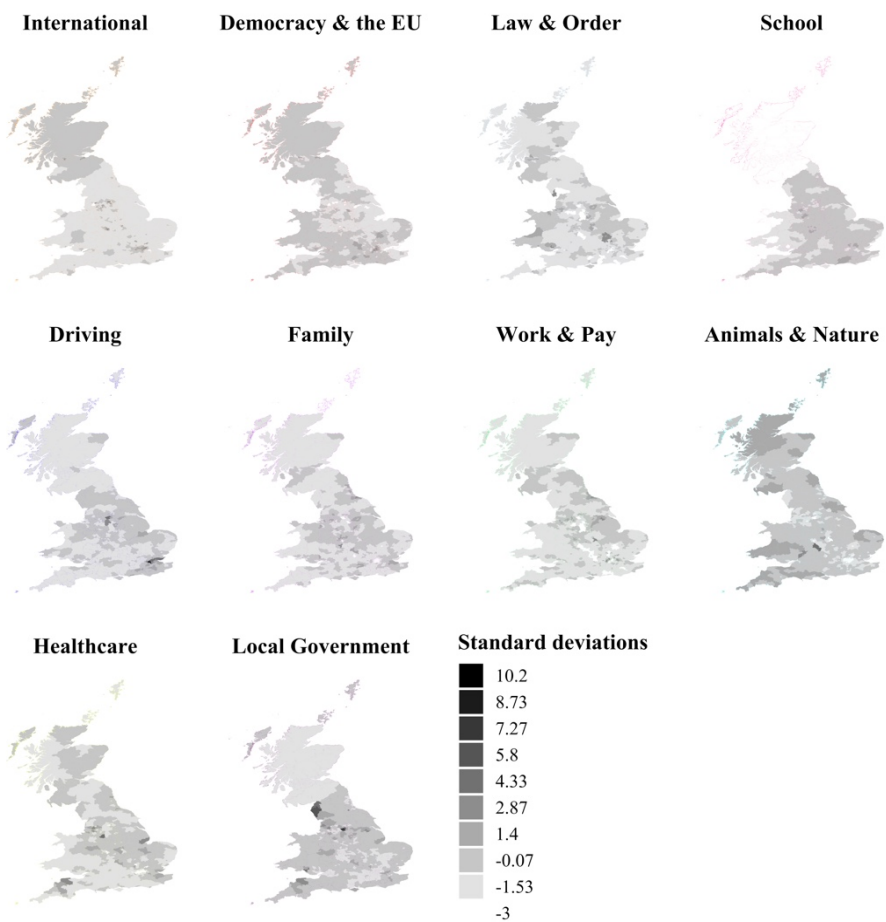

Figure 8: Issue maps. The prevalence of issues in each constituency. The darkness of the shading represents the number of standard deviations the percentage of signatures from each constituency for each issue are from the mean.

*Geographic clusters*

The results so far suggest a strong relationship between geography and petitions' issue. To further investigate this, we assign each constituency to one of 6 geographic clusters (hereon called 'Geo clusters'). We Cluster constituencies with Partition Around Medoids (PAM) clustering, based on the number of signatures given to each issue (see  ). The clustering shows distinct regions of petition signing, which complement the findings from the previous section. First, there is a clear divide between rural and urban constituencies. Geo clusters 1 and 2 are both heavily rural whilst Geo clusters 5 and 6 are primarily urban. Geo cluster 4 is a mix, comprising both rural and urban constituencies, all of which are located in the North East. Finally, there is a distinctive region which is mostly comprised of only Scotland (Geo cluster 3). The clusters are plotted in Figure 9.

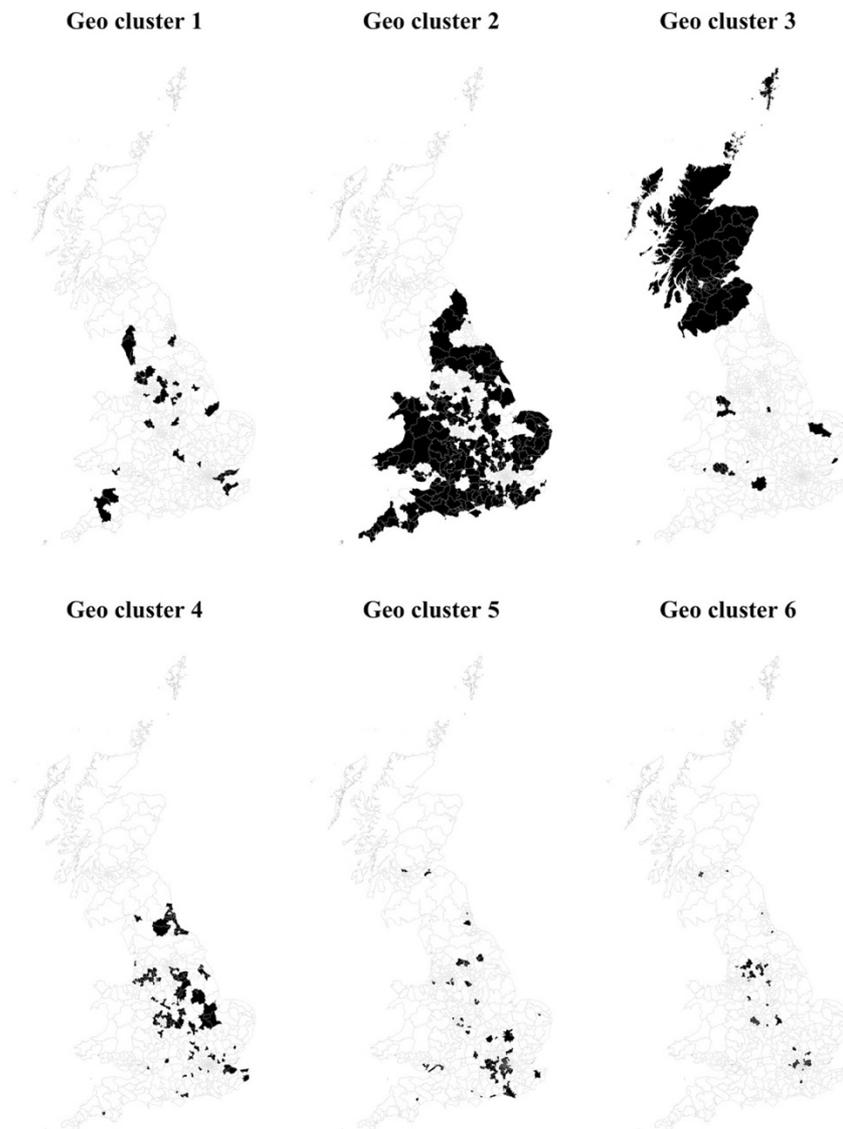

Figure 9: The constituencies assigned to the 6 geographic clusters based on the percentage of their signatures which are given to each issue.

The 6 Geo clusters have distinctive patterns of issue signing. Figure 10 shows the relative popularity of each issue for each Geo cluster by comparing the percentage of signatures. Geo clusters 1 and 2 (primarily urban constituencies) assign far more importance to the issues 'International' and 'Democracy & the EU' whilst Geo clusters 4 and 5 (primarily urban) favor 'Local Government' and 'Animals and Nature'. As noted in the previous section, Scotland (Geo cluster 3), ascribes considerably less importance to 'Local Government' and 'School'. This analysis indicates that the concerns of citizens (expressed through the issues of the petitions they sign) are linked powerfully to not only temporality, and the impact of exogenous events, but also geography. This opens up new avenues for research, including investigations of how geographic environment influences individuals' behavior and how geography can be a proxy measure for other issue-influencing factors, such as ethnicity, class and gender.

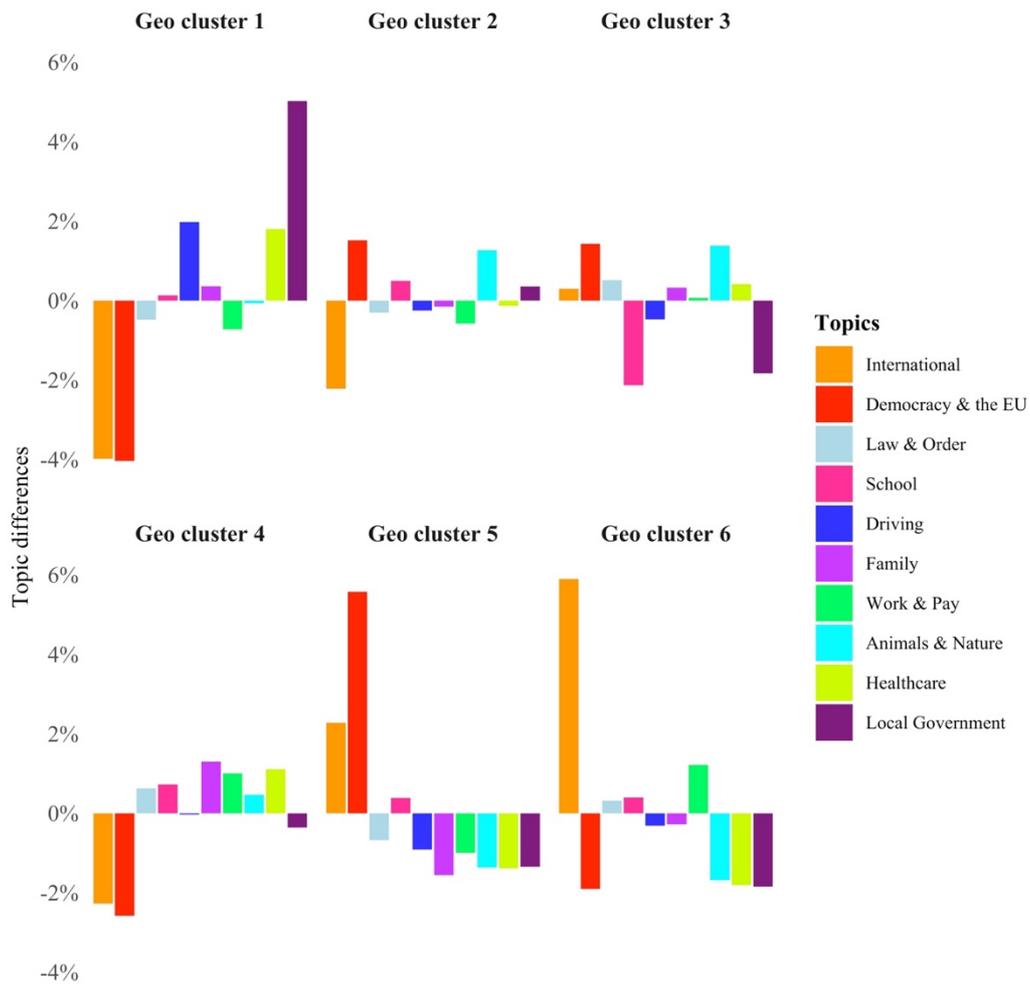

Figure 10: The popularity of each issue within each Geo cluster.

## 2.4 Comparison between the two approaches

Petitions are an excellent data source for understanding the concerns and priorities of citizens. They can be considered 'big data' as they contain large amounts of time-stamped granular transactional data and are available in real-time (Dumas et al. 2016). However, at present they are under-utilized in social scientific research and government services. To substantiate the claim that the body of petitions carry reliable signals we compare our ten issues with the results of Ipsos MORI's 'Issues Index' survey (Ipsos MORI 2019). Each month, Ipsos MORI asks a representative sample of ~1,000 British adults, 'What do you see as the most important issue facing Britain today?' Respondents can answer freely and do not have to choose from pre-established categories. The Issues Index dataset is widely used in the media (The Economist 2017) and political science research (Curtice 2017; Mellon 2013). During the 2015-2017 parliament (the period for which we have petitions data) 25 rounds of the survey took place. To compare this dataset with the issues extracted from the petitions data, we take the top ten issues from respondents during the period. The top ten Issues Index issues and top ten petition issues are shown in Table 3. Note that the petition issues 'Work & Pay' is associated with two of the Ipsos MORI Issues Index issues.

Table 3: Top ten Ipsos MORI Issues Index issues compared with the ten petition issues. The grey cells show the issues which have no direct equivalent in the other column.

| Ipsos MORI Issues Index | Petition issues |
|---|---|
| NHS/Healthcare | Healthcare |
| EU/Europe | Democracy & the EU |
| The Economy | Work & Pay |
| Education / Schools | School |
| Unemployment / Industry | Work & Pay |
| Defence / Foreign Affairs | International Affairs |
| Crime | Law & Order |
| Housing | Local Government |
| Poverty/Inequality | [No equivalent] |
| Immigration | [No equivalent] |
| [No equivalent] | Driving |
| [No equivalent] | Family |
| [No equivalent] | Animals & the Environment |

Table 3 shows that there is a close relationship between the issues that the public reports it is concerned by and the issues within petitions. Eight of the Issues Index issues relate closely to the petition iissues, and only 'Poverty/Inequality' and 'Immigration' are not explicitly included within the petition issues. Similarly, only three of the petition issues, 'Driving', 'Family' and 'Animals & the Environment', are not captured in the Issues Index. These discrepancies raise the question of why some issues are not identified through the survey methodology even though they are petitioned about, which is arguably a more expressive, time consuming and self-driven task. Potentially, such issues are not reported to interviewers because respondents do not perceive them as overtly 'Political' (Checkel and Katzenstein 2009). This analysis suggests that petitions could provide insight into political concerns which are not captured by traditional survey methodologies. Indeed, a key benefit of studying petitions is that they enable us to examine individuals' actual political concerns, as expressed via their behavior, rather than their stated concerns or intended actions (Margetts 2017). It is also likely that signing a petition is a sincerer expression by citizens of what concerns them as, once a petition is accepted onto the platform, there is no outside scrutiny. In contrast, surveys administered either through interviews or self-completed can suffer from many biases (Moy and Murphy 2016).

One important limitation of using online petitions to gain insight into the concerns of citizens is the well-noted 'digital divide' and evidence that Internet users are more likely to be younger, white, and well-educated than the general population (Blank 2017; van Deursen and van Dijk 2014; Friemel 2016). However, there is some preliminary evidence that a broad range of the population sign petitions. For instance, research by Melo and Stockemer finds that across France, Germany and the UK the propensity to sign petitions is relatively stable, albeit weakly curvilinear, with age; adults (those between 34 and 64) have a 35.5% chance of signing a petition, compared with 31.4% for young adults (< 34) and 24.6% for the elderly (> 64) (Melo and Stockemer 2014). Similarly, in the 2013 Oxford Internet Survey five sub-cultures of Internet users were identified, based on attitudes towards Internet use, and all five sub-cultures reported signing petitions, from 6% for the 'e-Mersives' to 17% for the 'Adigitals' (Dutton and Blank 2013). Thus, whilst petitions should not necessarily be considered representative of the views of the entire public, there is evidence that their use is widespread and not restricted to just one particular sub-group.

# 3 Discussion and Conclusion

In this paper, we first provided an overview of the number of signatures for each petitions, and showed the impact of the governments' response thresholds on the number of signatures petitions receive. Then, we identified ten issues and showed their overall prevalence, their relationship with the probability of a petition being successful and how they relate to each other. In the next step, we studied the temporal dynamics of issues and showed that these vary considerably across issues. We presented a measure of entropic change which can be used to identify periods of intense volatility. Then, we analyzed the geographic distribution of issues. We showed systematic geographic variations in the importance of the ten issues and identified 6 distinctive geo clusters. Finally, we argued that petitions should be better used in social scientific research, and provided evidence that they are a powerful way of understanding the concerns and priorities of citizens in a timely manner by comparing them with survey data from Ipsos MORI.

This research demonstrates how the thematic content of petitions can be analyzed in order to understand issues which concern the UK public. It also shows that the UK public's interest in issues is complex and heterogeneous: there are important geographic and temporal dynamics which need to be taken into account. Our work serves an important democratic function by increasing the democratic voice of citizens. In our method, every signature given to every petition is analyzed systematically and in-depth, in a computationally efficient manner. This ensures that all of these 'micro-acts' of participation are given due attention and that not only hyper successful petitions are listened to. Our method, which uses LDA topic modelling, is reproducible and could be applied in near-real time to provide MPs with insights about the concerns of their constituents. In the future, the work can be extended by integrating analysis of petitions' content with their sentiment and ideological stance. This would provide greater insight into how constituents view issues and how they want them addressed. Overall, we believe that petitions are a promising area for future research as they are organically created by citizens, and as such have made all of our code and data publicly available.

# 4 Data and Methods

We collect all petitions submitted to the UK Government during the 2015-2017 parliament (available at https://petition.parliament.uk) using a script written in Python. 31,731 petitions were submitted, of which 10,950 were accepted onto the platform and 20,781 rejected. We retain only the 10,950 accepted petitions. 31,473,493 signatures are made during the period of which 1,052,510 are from citizens in countries outside of the UK (3.3%). We remove these from the dataset, leaving 30,420,983 signatures.

To make a petition, users have to outline a proposed action and provide some background information, and have the additional option of providing further details. We combine these three free-text entry fields into a single variable for each petition. Initially, there are 33,115 unique terms in the corpus, and each petition has on average 81 non-unique terms. We clean the text by transforming it to lower case, removing punctuation, removing numbers, removing stopwords, stripping whitespace and stemming words using the Porter stemming algorithm (Porter 1980). A Document-Term Matrix is created from the corpus of cleaned text. To reduce the chance of over-fitting on infrequent terms, we reduce sparsity by removing terms which appear in less than 0.1% of all petitions. This reduces the average number of non-unique terms in each petition from 55 to 44, and the number of unique terms in the entire corpus from 33,115 to 3,592.

## 4.1 Topic Modelling

LDA topic modelling is a well-established multi-membership NLP method for extracting the themes in a corpus of text without the need detailed qualitative reading (Grimmer and Stewart 2013). Topic models use the observable variables in a corpus – (i) documents (in our case each petition is a document), and (ii) words – to model the unobserved or 'latent' variables, such as the topics (Blei et al. 2003). Topic models are fitted by estimating the latent variables (the multinomial topic distributions over words and theta values) through sampling and expectation-maximization (Steyvers and Griffiths 2004). It is a mixed-membership model, which means that petitions are assumed to contain a mixture of all topics, rather than being assigned to just one. The same applies to the words, i.e., they are assigned to multiple topics, although with different weights. Effectively, when implemented, topic models move backwards through the generative process described here to uncover the topics which led to the observed words in the documents. One aspect of this generative process is that word order is not directly modelled; accordingly, topic models only use a simplifying 'Bag of Words' assumption.

Topic models contain three hyperparameters which determine the fit of the model: K (the number of topics), alpha (the distribution of topics within documents), and beta (the distribution of words within topics). Lower values of alpha increase the probability that a value of theta will be selected which is skewed towards a few dominant topics. Lower values of beta increase the probability that for the topics (which are multinomial distributions over words) higher probabilities are assigned to the most likely words. This means that topics are primarily composed of just a few dominant words. As a result, it can be easier to separate topics, which is often appropriate when the topics concern very different subjects or when the number of topics is low. We test for different values of K, beta and alpha, and fit K = 10, alpha = 0.1, beta = 0.1.

A well-recognized problem in the literature on topic modelling is how to label topics accurately, reproducibly and quickly (Chang et al. 2009; Lau et al. 2011). If a non-domain expert is used to annotate the topics then they can easily misinterpret them. Through discussion between the authors of this paper, we agree on labels for all ten topics based on our expertise in the field of political science as well as previous research on petitions. All of our topics are identifiable as substantive political issues, and all of the topics are mutually exclusive (in that they pertain to clearly distinguishable issues), which is an unexpected positive result. The ten topics we identify are shown above in the Results section.

To validate topic coherence we use the qualitative extrinsic method of 'word intrusion', as described by Chang et al.. This method is appropriate as it measures 'how semantically "cohesive" the topics inferred by a model are and tests whether topics correspond to natural groupings for humans' (Chang et al. 2009, 2). Subjects are presented with six randomly ordered words, one of which is *not* associated with the topic with a high probability. The subjects need to identify the word which does not belong with the others; the *intruder*. For coherent topics it should be easy to identify the intruding word. In their empirical analysis, Chang et al. suggest that values above 75% indicate good topic coherence.

To complete this task, we use three students as subjects who are not otherwise affiliated with the research. For each of the ten topics we show them six words; the five most probable words from the topics and then one word randomly selected from a pool of words with low probability in the topic (defined as words with a probability less than or equal to the median probability). The six words are shuffled before being presented to the subject. Overall, we achieve accuracy of 86.4%, and for all topics accuracy is greater than 66% (indicating 2 of 3 subjects identified the intruding word correctly). This is evidence of the robustness of our topics.

Table 4: Results of word intrusions tests to measure topic coherence

| Topic Name | Top five words | Three randomly selected words | Percentage accuracy |
|---|---|---|---|
| International affairs | British, Govern, Country, Nation, World | License, Employee, Homosexual | 100% |
| Democracy & the EU | Vote, Referendum, Govern, Parliament, Leave | Language, Engine, Holocaust | 100% |
| Law & Order | Law, Police, Act, Public, Protect | Lifetime, Enter, Hat | 66% |
| School | School, Children, Student, Education, Year | Lisbon, Error, Impair | 100% |
| Driving | Road, Car, Use, Driver, Drive | Leaflet, Epidemiology, Hotel | 100% |
| Family | Children, Child, Parent, People, Family | License, Entity, Ideology | 100% |
| Work & Pay | Pay, Tax, Work, Year, Cost | Language, Energy, Hostility | 100% |
| Animals & the Environment | Dog, Animal, Ban, Use, Food | Libya, Equipment, Implemented | 66% |
| Healthcare | NHS, Health, People, Care, Mental | Likely, Exact, Implementation | 66% |
| Local Government | Govern, Housing, Local, Will, Council | Leaflet, Emotion, Horror | 66% |

We check the validity of the topic model by randomly selecting 10 petitions for each topic (total n = 100) which have a high probability (p > 0.95) and check whether their topic assignment is correct. We find that 89 of the 100 petitions are assigned correctly, which suggests that our method performs well.

### 4.2 Geographic clustering

We assign constituencies to Geo clusters. To do so. we calculate the number of signatures given to each issue within each constituency. We then transform this into a Z-score, based on the distribution of signatures from each constituency. This lets us capture the relative importance assigned to the issue. On this transformed data, we implement Partition Around Medoids (PAM) clustering, which assigns each constituency to only one Geo cluster. PAM is a single-membership clustering method, which is similar to k-means. It assigns instances to clusters by minimizing a sum of dissimilarities instead of a sum of squared Euclidean distances (as with k-means) (Xu and Wunsch 2008). We use 6 clusters after fit analysis indicates a range of 5 to 10 is most suitable.